# Magnetic field induced augmented thermal conduction phenomenon in magneto-nanocolloids

**Ajay Katiyar[a,b,*], Purbarun Dhar[b, #], Tandra Nandi[c,×], Sarit K. Das[b, e,$]**

[a] Research and Innovation Centre (DRDO), Indian Institute of Technology Madras Research Park, Chennai–600 113, India (1st affiliation of the author)
[b] Department of Mechanical Engineering, Indian Institute of Technology Madras, Chennai–600 036, India
[c] Defence Materials and Stores Research and Development Establishment (DRDO), G.T. Road, Kanpur–208 013, India

*Corresponding author: Electronic mail: ajay_cim@rediffmail.com
*Tel No: 044-22548-212
*Fax: 044-22548-215
$Corresponding author: Electronic mail: skdas@iitm.ac.in
$Phone: +91-44-2257 4655
$Fax: +91-44-2257 4650
# E-mail: pdhar1990@gmail.com
× Email: tandra_n@rediffmail.com

[e] School of Mechanical, Materials and Energy Engineering (SMMEE), Indian Institute of Technology Ropar, Rupnagar–140001, India (Present address of author)

# Abstract

Magnetic field induced drastically augmented thermal conductivity of magneto-nanocolloids involving magnetic oxide nanoparticles, *viz.* $Fe_2O_3$, $Fe_3O_4$, Nickel oxide (NiO), Cobalt oxide ($Co_3O_4$), dispersed in different base fluids (heat transfer oil, kerosene, and ethylene glycol) have been reported. Experiments reveal the augmented thermal transport under the external applied magnetic field, with kerosene based MNCs showing at relatively low magnetic field intensities as compared to the heat transfer oil and EG based MNCs. A maximum thermal conductivity enhancement ~ 114 % is attained at 7.0 vol. % concentration and 0.1 T field intensity for $Fe_3O_4$/EG magneto-nanocolloid. However, a maximum ~ 82% thermal conductivity enhancement is observed for $Fe_3O_4$/Kerosene magneto-nanocolloid for the same concentration but relatively at low magnetic field (~ 600 G). Thereby, a strong effect of fluid as well as particle physical properties on the chain formation propensity, leading to enhanced conduction, in such systems is observed. $Co_3O_4$ nanoparticles show insignificant effect on the thermal conductivity enhancement of MNCs due to their minimal magnetic moment. An analytical approach has been proposed to understand the mechanism and physics behind the thermal conductivity enhancement under external applied magnetic field, in tune with near field magnetostatic interactions as well as Neel relaxivity of the magnetic nanoparticles. Furthermore, the analytical model is able to predict the phenomenon of enhanced thermal conductivity as a function of physical parameters such as chain length, size and types of nanoparticles, fluid characteristics, magnetic field intensity, saturation magnetic moment, nanoparticle concentration etc. and good agreement with the experimental results has been observed. The present study, involving analytical as well as detailed experimental analyses of an important phenomenon of thermal conductivity enhancement can find utility in design of 'smart fluids' for thermal management of devices with high thermal loads.

*Abbreviations*

MNCs - Magneto-nanocolloids

HTO - Heat Transfer Oil

EG - Ethylene Glycol

DI – De-ionized water

TEM - Transmission Electron Microscopy

HRSEM - High Resolution Scanning Electron Microscopy

DLS - Dynamic Light Scattering

SAED - Selected Area Diffraction Electron Diffraction

VSM - Vibrating sample magnetometer

CMF - Critical Magnetic Field

NPs - Nanoparticles

MFI – Magnetic Field Induced

# 1. Introduction

Enhancement of transport phenomena caliber in nanofluids (comprising homogeneously dispersed NPs in conventional fluids) and its application in variant fields such as smart cooling of electronic devices and automotive components, vibration damping, nano-finishing, cancer treatment and drug delivery, sensors, switches etc. has attracted the attention of the scientific and engineering community for the past decade. Generally, scientific literature on transport properties of ferrocolloids reports the enhancement of thermal conductivity of carrier fluids [1-7] upon seeding with thermally conducting magnetic NPs. However, very few literature reports the augmented thermal conductivity of ferrocolloids under the influence of magnetic field [2, 6, 7]. Thermal conductivity being a significant property of smart fluids in heat transfer applications has led several researchers over the years to study thermal conductivity caliber of nanofluids [8-12]. However, the levels of enhancement exploitable from such colloidal systems can be further augmented since magnetic nanofluids exhibit enhanced thermal conductivity when subjected to an external magnetic field [13] due to formation of long and stable chains along the magnetic field [14] through which the heat wave 'short-circuits'. The gap between NPs decreases with increasing volume fraction as well as decreased particle size; thereby increasing the strength of particle- particle interaction on the application of externally applied field due to initiation of field induced Kelvin body forces [15], leading to higher effective thermal transport.

The thermal conductivity of nanofluids increases with particle concentration regardless the type of dispersed phase [16-25], given the higher thermal conductivity of solid as compared to fluids. The orientation of the magnetic field with respect to temperature gradient also influences the thermal transport property of the magnetic fluids under the influence of external applied magnetic field and insignificant effect was observed if the orientation of the magnetic field is perpendicular to the applied temperature gradient [6]. Thus, thermal properties of the smart magnetic nanofluids can be tuned by controlling the nature and strength of the externally applied magnetic field [26, 27]. However, the effect of thermal transport in magnetic nanofluids vis-à-vis magnetic field intensity, type of magnetic particles, particle size and concentration and fluid features and the underlying mechanisms are yet to be fully comprehended. The current article presents detailed experimental analysis on the augmented thermal conductivity of different MNCs formulated by uniformly dispersing $Fe_2O_3$, $Fe_3O_4$, NiO, and $Co_3O_4$ NPs in different base fluids such as heat transfer oil, kerosene and ethylene glycol (EG) under the influence of magnetic fields. In the present study, NPs concentration and magnetic field have been varied from 0.0 – 7.0 vol. % and 0.0 – 0.2 T, respectively. The present work also proposes an analytical formulation which can accurately predict the system physics as well as the magnitude of enhanced thermal conductivity based on system parameters such as concentration, particle size and nature, fluid characteristics and field strength. The model has been observed to predict the experimental results with appreciable accuracy.

## 2. Materials and methodology

Cobalt oxide and Iron (III) oxide ($Fe_2O_3$) NPs utilized in the present study have been procured from Nanoshel Inc. (USA) and Alfa Aesar (USA) respectively and dried at 100 °C for 2 h before use. Ethylene glycol has been procured from sigma Aldrich (purity ~ 99.5%, anhydrous), kerosene (laboratory grade) has been procured from Sisco Research Labs (India) and heat transfer oil utilized is Therminol® 55. The $Fe_3O_4$ and NiO particles have been chemically synthesized (discussed in the next section).

## 2.1. Synthesis of nanomaterials

### 2.1.1. $Fe_3O_4$ NPs

Iron (II, III) oxide ($Fe_3O_4$) NPs has been synthesized by co-precipitation technique based on protocol reported previously by the present authors [28]. Ferric chloride hexahydrate ($FeCl_3.6H_2O$, Merck) and ferrous chloride tetrahydrate ($FeCl_2.4H_2O$, Merck) act as the source of iron oxide and 30 % ammonia solution (Sisco Research Labs) acts as the reducing agent. The iron salts are individually dissolved in DI water to produce 1M stock solutions. The solutions are mixed in a volumetric ratio of $Fe_3^+$: $Fe_2^+$ = 2:1 and continuously stirred at 3000 rpm with a mechanical stirrer while heating to a temperature of 75–85 °C. The ammonia solution is added drop wise to the solution while stirring until the solution changes its color to a dark brownish black and the pH of 11-12 is attained. The solution is allowed to stand overnight while heating at 50 °C for the ammonia to vaporize out and then centrifuged at 10000 rpm to obtain the precipitate. It is then washed with DI water several times, until the pH becomes ~ 7.0. The particles are then washed with Acetone (Merck) to remove organic/inorganic impurities and dried in inert atmosphere for a day to obtain the dry NPs.

### 2.1.2. NiO NPs

NiO NPs have been synthesized via thermal decomposition of nickel hydroxide. Initially, a 0.1 M stock is prepared by dissolving nickel chloride hexahydrate ($NiCl_2.6H_2O$, Sigma Aldrich) in ethanol (absolute, Sisco Research Labs). The solution is added to a hydrazine monohydrate solution (molar ratio 5) while stirring at room temperature. The pH of the solution was elevated to 12.0 using finely powdered sodium hydroxide (NaOH). The solution is stirrer over 2-3 hours and a dark greenish coloration indicates the formation of nickel hydroxide [$Ni(OH)_2.0.5H_2O$]. The precipitate is collected via centrifugation and washed repeatedly with DI water and acetone to remove traces of the reactants. Subsequently, the precipitate is dried in an inert oven over a temperature range of 500-600 °C to transform the nickel hydroxide NPs to nickel oxide NPs via thermal decomposition. The particles are collected and washed with acetone to remove any further residues and dried in an oven at 80°C overnight. The process chemical reaction for the same is mentioned below:

$$NiCl_2 \cdot 6H_2O + 6C_2H_5OH \longrightarrow [Ni(C_2H_5OH)_6]Cl_2$$
$$[Ni(C_2H_5OH)_6]Cl_2 + N_2H_2 \longrightarrow 6C_2H_5OH + [Ni(N_2H_4)_n]Cl_2$$
$$[Ni(N_2H_4)_n]Cl_2 + 2NaOH \longrightarrow nN_2H_4 + Ni(OH)_2$$
$$Ni(OH)_2 \xrightarrow{500°C-600°C} NiO$$

## 2.2 Characterization of NPs

Size and morphology of NiO and $Co_3O_4$ NPs have been confirmed by TEM, whereas, that for $Fe_2O_3$ and $Fe_3O_4$ NPs by HRSEM. Samples for obtaining HRSEM images of the particles are made by dispersing very small quantities of $Fe_2O_3$ and $Fe_3O_4$ NPs in pure ethanol, followed by ultrasonication for 15 minutes. Subsequently, a drop of the solutions are taken on the corresponding stubs individually and kept in an oven at 65 °C for 2-3 hours for complete evaporation of the ethanol. The HRSEM images of $Fe_2O_3$ and $Fe_3O_4$ NPs are illustrated in Fig. 1 (a) and (b) respectively. The dimensions of $Fe_2O_3$ and $Fe_3O_4$ nanoparticle have been observed to exist over the ranges of 20 - 30 nm and 60 - 90 nm respectively and shapes are confirmed as near spherical and pyramidal (inverse spinel structure) respectively. The NiO and $Co_3O_4$ NPs are imaged by TEM utilizing carbon coated copper grids with 300 mesh size. The grid preparation for TEM images is done in the following fashion; initially, particles are dispersed in ethanol individually by utilizing a probe type ultrasonicator and consequently, individual drop of samples are taken on the grids. The excess fluid is kept for draining and samples are dried before any measurements. Figure 1(c1) and (d1) illustrate the TEM images of NiO and $Co_3O_4$ NPs. The TEM images demonstrate that NiO and $Co_3O_4$ NPs are polyhedral (in general of hexagonal cross section) and oblate in shape with mean diameter of ~ 35 nm and ~ 25 nm, respectively. Figures 1 (c2) and (d2) reveal the SAED pattern of NiO and $Co_3O_4$ NPs respectively with the bright spots arranged in concentric patterns confirming the crystalline nature of the NPs, arising from Bragg diffraction from each crystallite structure. Figures 1(c3) and (d3) show the DLS plots for nickel and cobalt oxide NPs respectively with the sharp peaks indicating the most probable particle sizes. The present analysis reveals NiO and $Co_3O_4$ NPs as 40 nm and 30 nm with very narrow scattering in particle size, respectively.

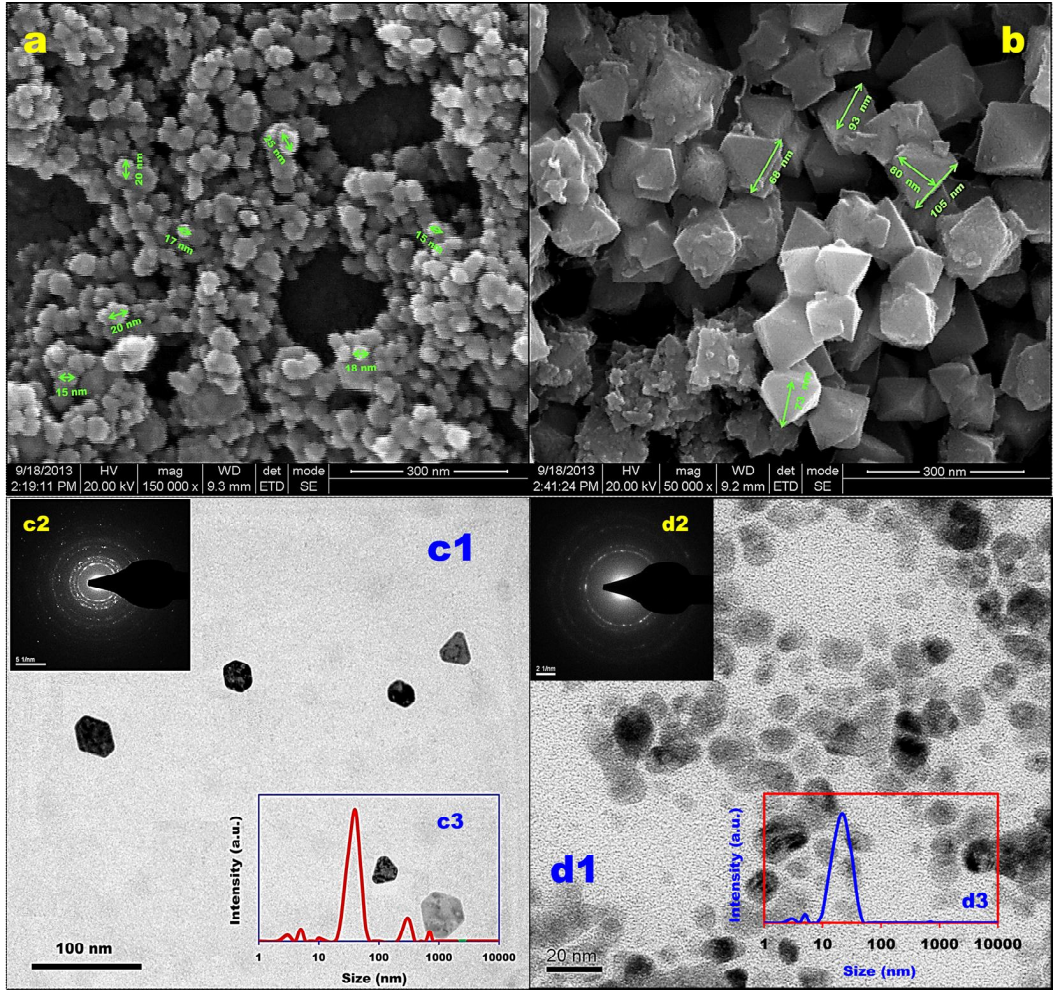

**Figure 1:** HRSEM images of particles. (a) Reveals average size of $Fe_2O_3$ NPs over a range of 20 - 30 nm with almost spherical shape. (b) Indicates the average size of $Fe_3O_4$ particles over a range of 60 - 90 nm with pyramidal structure. (c1) TEM image of NiO particles reveals average dimension over a range of 30 - 40 nm. Inset image (c2) SAED pattern of NiO NPs reveal crystalline structure and (c3) DLS based characterization of the sample, showing size distribution of the dispersed NiO NPs. (d1) TEM image of $Co_3O_4$ NPs shows average particle diameter over a range of 20 - 30 nm. Inset images (d2) and (d3) illustrate the SAED and DLS characterization, respectively for the $Co_3O_4$ NPs.

The magnetic characteristics of the NPs have been measured by a VSM at 300 K (corresponding to the experimental conditions) and the magnetization curves (M vs. H) for the same are illustrated in Fig. 2. It is observed that at 300 K, $Fe_3O_4$, $Fe_2O_3$, NiO and $Co_3O_4$ NPs have magnitude of magnetic saturation as 56, 36, 4.6 and 1.7emu/g, respectively. It is observed from the measurements that the NPs demonstrate no magnetic hysteresis, confirming super paramagnetic behaviour. This is further confirmed by the fact that the particle sizes are smaller in magnitude than the critical domain size for the corresponding magnetic material, e.g. the critical domain size for iron (II,III) oxide is ~ 120 nm and particles with diameters smaller than this exhibit superparamagnetism.

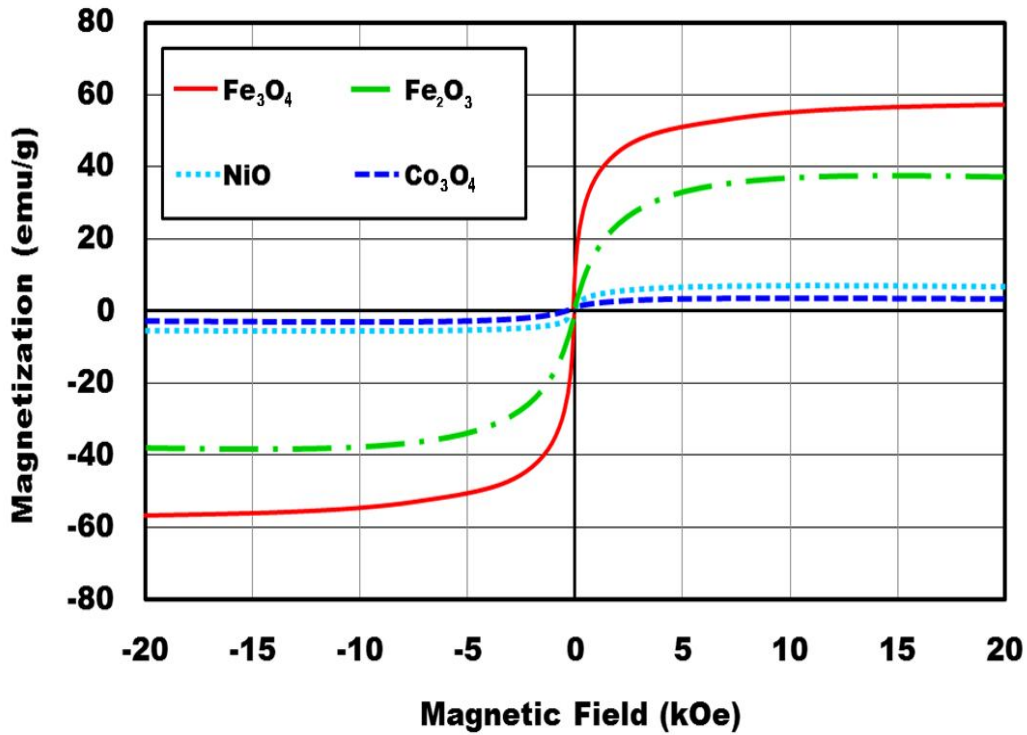

**Figure 2:** Magnetization curves for different NPs at 300 K.

## 2.3. Synthesis of Magneto-nanocolloids

MNCs have been synthesized in accordance to protocol reported by the present authors [29]. Initially, the prerequisite amount of different NPs such as $Fe_3O_4$, NiO, $Fe_2O_3$, and $Co_3O_4$ NPs is homogeneously dispersed within the base fluids over a volumetric concentration range of 0 – 7 vol. %. Subsequently, the mixture is subjected to mechanical stirring for twenty minutes and further sonicated for 2 - 3 hrs at 65% amplitude and 10:10 pulse rate utilizing a probe type sonicator (VCX-750 Sonics & Materials, USA) to disrupt any agglomeration of NPs. A small amount of oleic acid is used as capping agent to improve the long term stability of the oil based colloids whereas no surfactant is required for the EG based MNCs due the high viscosity of the fluid. It is observed that $Fe_2O_3$ based MNCs are more stable due to small size of NPs and the stability is retained even in the presence of strong magnetic field intensities. Images of representative samples of $Fe_2O_3$/EG, NiO/kerosene and $Fe_3O_4$/HTO MNCs have been illustrated in Fig. 3 (b), (c), and (d), respectively. (All synthesized colloidal samples retained the color equivalent to that of dispersed particles).

## 3. Instrumentation

The magnetic field applied across the colloidal samples is generated by a custom made electromagnet (EM-100, Polytronic Corpn., India) and which can vary the field intensity over a range of 0 – 1.2 T for pole spacing of ~ 10 mm. The setup consists of a DC power supply (rated 5A, 60V) to control the current or voltage across the electromagnet pole windings. The device consists of digitized voltmeter and ammeter to monitor output voltage and current. The excitation coils are mounted on two poles and fitted with two vertical yoke members of electromagnet. Adjustable flat face cylindrical iron billets of diameter 100 mm act as the pole shoes. Since large heat is generated by the coils during operation, a cooling circuit employing water is utilized to cool the coil mounts. The gap between the poles is adjustable over a range of 0 – 100 mm with a two way knobbed wheel screw adjusting system. A Gauss meter is employed to measure the magnetic field intensity between the two poles and used to calibrate it against the different current or voltage applied to the windings. The system is calibrated by keeping the gap equivalent to the diameter of the test vessel.

Thermal conductivity of the MNCs is measured by KD2 Pro thermal property analyzer (Decagon Devices Inc. USA) working on the principle of transient hot wire technique. Thermal conductivity is computed by the device utilizing the change in temperature along the infinitely long and thin line source as a function of time period in which the change occurs. The equation for same can be expressed as (eqn.1):

$$k = \frac{q(\ln t_2 - \ln t_1)}{4\pi(\ln \Delta T_2 - \ln \Delta T_1)} \qquad (1)$$

Where '$q$' indicates the constant heat flux supplied to line source, '$\Delta T_1$' and '$\Delta T_2$' the change in temperature at time '$t_1$' and '$t_2$' respectively. The probe is calibrated using standard fluid samples supplied by the manufacturer and can be used to measure the thermal conductivity of fluids over a range of 0.02 – 4.0 W/mK. The uncertainty in thermal conductivity measurements was found to be less than ±2 %. The arrangement for thermal conductivity measurement is made in such a way that the temperature gradient generated and the magnetic field are always parallel since no enhancement in the thermal conductivity is observed for perpendicular arrangement of the temperature gradient with respect to the magnetic field [27]. The complete experimental setup has been illustrated in Fig. 3(a).

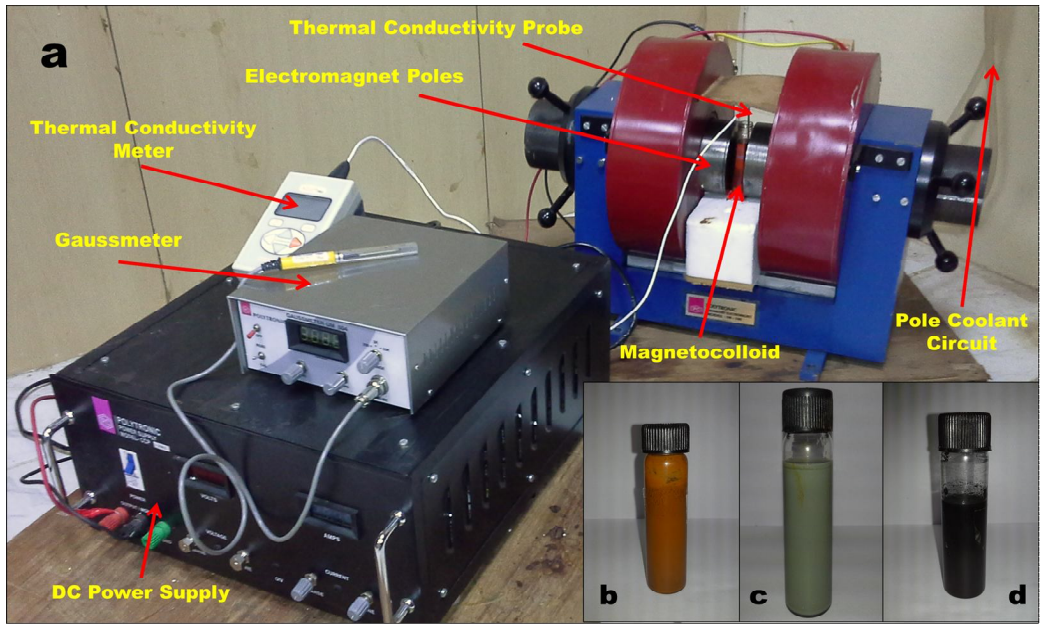

**Figure 3:** (a) Experimental setup for thermal conductivity measurement in presence of applied magnetic field. (b) $Fe_2O_3$/EG MNC with rust color (c) NiO/kerosene MNC with as a green color (d) $Fe_3O_4$/HTO MNC with black color.

# 4. Results and Discussions

## 4.1 Fe$_3$O$_4$ magneto-nanocolloids

Experiments reveal that the thermal conductivity of MNCs increases as function of the applied magnetic field intensity. Fig. 4 (a) illustrates the thermal conductivity enhancement of the Fe$_3$O$_4$/EG MNCs with increasing magnetic field intensity. The maximum rise in the MFI thermal conductivity is observed at ~ 0.1 T and consequently the thermal conductivity decreases after this CMF. The magnitude of maximum thermal conductivity for the family of Fe$_3$O$_4$/EG MNCs is observed to be as 0.64 W/mK at the CMF for a concentration of 7 vol. %. The magnitudes of thermal conductivity and percentage enhancement with respect base colloid in absence of field have been illustrated in Fig. 4 (b) and (c) respectively. A maximum ~ 114% enhancement in the thermal conductivity with respect to the base colloid is observed at 0.1 T and 7 vol. % concentration. The magnitude of thermal conductivity enhancement for the Fe$_3$O$_4$/HTO MNCs as a function of magnetic field is illustrated in Fig. 5 (a) and the corresponding enhancement of thermal conductivity with respect to the base colloid is illustrated in Fig. 5 (b). It is observed that a maximum thermal conductivity of 0.365 W/mK and corresponding enhancement of ~ 110 % is obtained at 0.08 T and 7 vol. %. The magnitude of maximum thermal conductivity for Fe$_3$O$_4$/kerosene MNCs is 0.4 W/mK (illustrated in Fig. 6 (a)) and the corresponding enhancement is ~ 100 % (Fig. 4.3 (b)) at 0.06 T and 7 vol. %.

The phenomenon of enhancement of thermal conductivity can be explained on the basis of chain formation or fibrillation due to particle-particle interactions and alignment along the magnetic field lines [14], leading to formation of conducting pathways for heat to diffuse along. The NPs remain in a randomly distributed state throughout the colloid in the absence of magnetic field (as illustrated qualitatively in Fig. 7 (a)). When the MNCs are subjected to magnetic fields, the NPs, by virtue of their magnetic moment, align along the magnetic field lines to form chained structure or fibrils (as illustrated in Fig. 7 (b)). However, thermal conductivity enhancement also depends on the strength and orientation of the chains. If the particles are loosely packed within the fibrils, such as occurring at low magnetic fields or low concentrations, the effective volume of the fluid in between two consecutive particles within the chains is higher and thereby the chain's ability to conduct heat is low. As the magnetic field intensity increases, the force of attraction among the NPs is also increased and reaches maxima at CMF. The maximum magnetic field sustained by the MNCs depends on the magnetic saturation of NPs, the particle size and the viscosity of the base fluid. The packing density of NPs is high at CMF for a specific MNC along the chains and force of attraction among the chains is balanced by the inter-particle magnetic force and consequently, the strength of formed chains is high, resulting in high thermal conductivity of the MNCs.

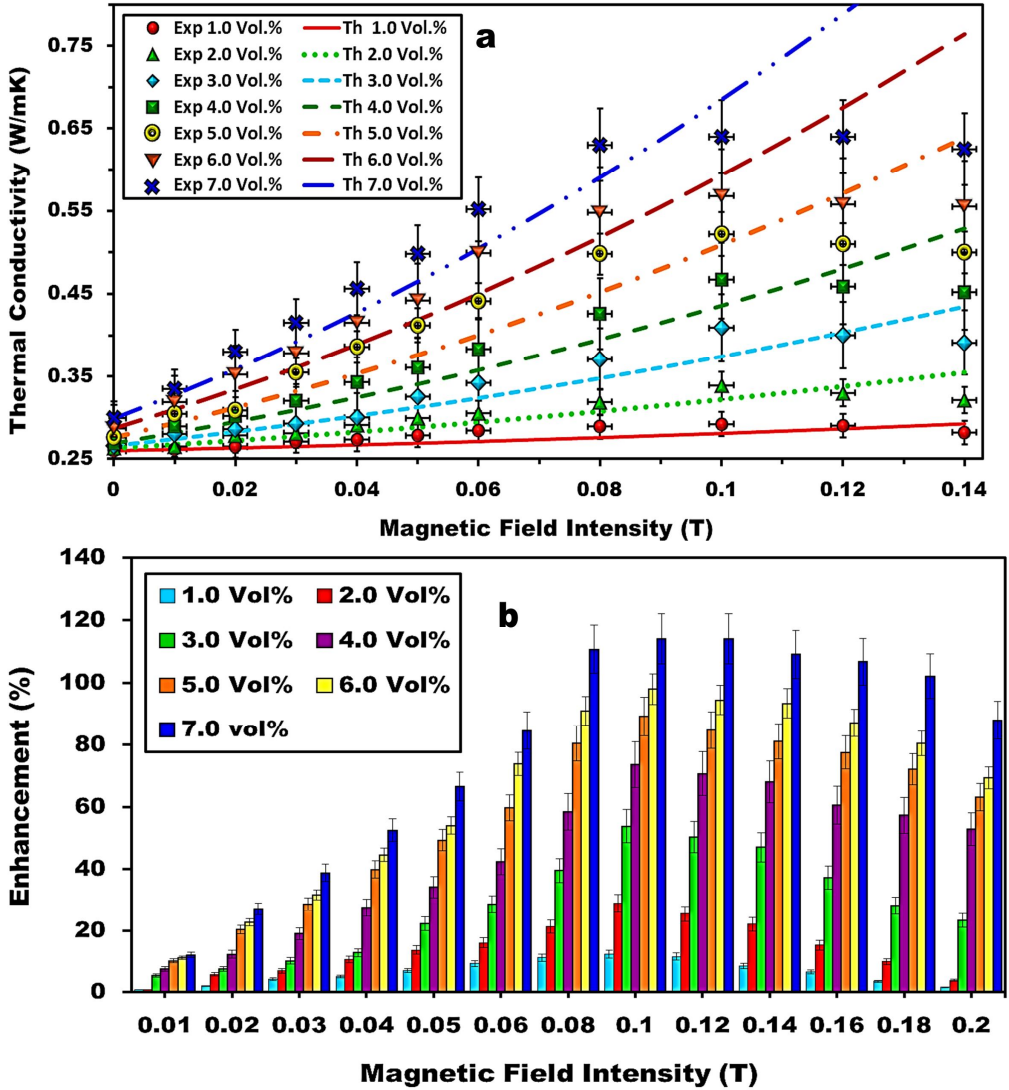

**Figure 4:** Thermal conductivity enhancements of $Fe_3O_4$/EG as function of magnetic field intensity. (a) The magnitude of thermal conductivity and validation of the analytical model with respect to experimental results. (b) The magnitude of enhancements of thermal conductivity with respect to base colloid at different NP concentrations and field intensities.

It has been observed that the CMF shifts towards lower magnitudes of field intensity with change in the base fluid from EG to HTO to kerosene. This phenomenon can be described based on the viscosity of base fluids. Intuition tells that in a fluid with lower viscosity, the dispersed particles have less resistance or hindrance against migrating along the applied field lines and getting aligned along the same and forming chained structures, which act as the pathways of enhanced heat conduction. As the viscosity of base fluid increases from kerosene to HTO to EG, the particles experience higher resistance towards fibrillation. Accordingly, chains form at higher values of magnetic field. However, it is also noteworthy that enhanced viscosity also dampens out thermal fluctuations or Brownian randomness to large extents, thereby leading to the formation of chains with higher thermal stability, leading to higher values of augmented conductivity in the EG based MNCs.

Beyond the CMF, zippering or lateral coalescence of the chains occur and leads to reduced heat conduction caliber [30]. Below and up to the CMF for a particular MNC, the field ensures that the chains are in general of equal length and as they laterally approach each other, the magnetic repulsion between them ensures equitable distribution of the chains within the colloid [31]. However, as the field surpasses the CMF, the field is strong enough for the colloid under question to induce inequality in chain lengths, as the particles closer to the core of the flux lines are more susceptible to form longer chains than those under the influence of the field lines away from the flux core. Consequently, when the chains have different lengths, the probability of zippering of chains enhances due to mismatch of repulsive strengths and dominance of attraction by the longer chains composed of more individual particles. Furthermore, the zipped state is one that the system tries to attain as the lowest energy state of the system consists of the chains clusters in the zipped fashion rather than the existence of individual, free standing chains [32].

The head to tail aggregation of individual particles, which is essential to form single chain, does not involve any significant energy barrier. However, the once aligned, the particles possess interaction energies with which they interact with the neighboring particles, leading to increment of the system energy in the chained configuration. Thereby, the chains interact among themselves kinetically to reach the total lowest energy state [30]. The dipole moment per unit length induced within the magnetically aligned chains interacts with of lateral field induced by chain fluctuations [33], essentially reducing the mean chain spacing. Moreover, the magnitude of the interaction energy, either attractive or repulsive in nature, depends on the magnetic field intensity (H), the separation distance ($d_{sep}$) between the chains and the thermal state of the system. The interaction energy per unit length (E) is expressible as (Eqn. 2) [34]

$$E \approx \frac{\chi H (\mu_0 k_B T)^{1/2} a^{5/2}}{(d_{sep})^2} \qquad (2)$$

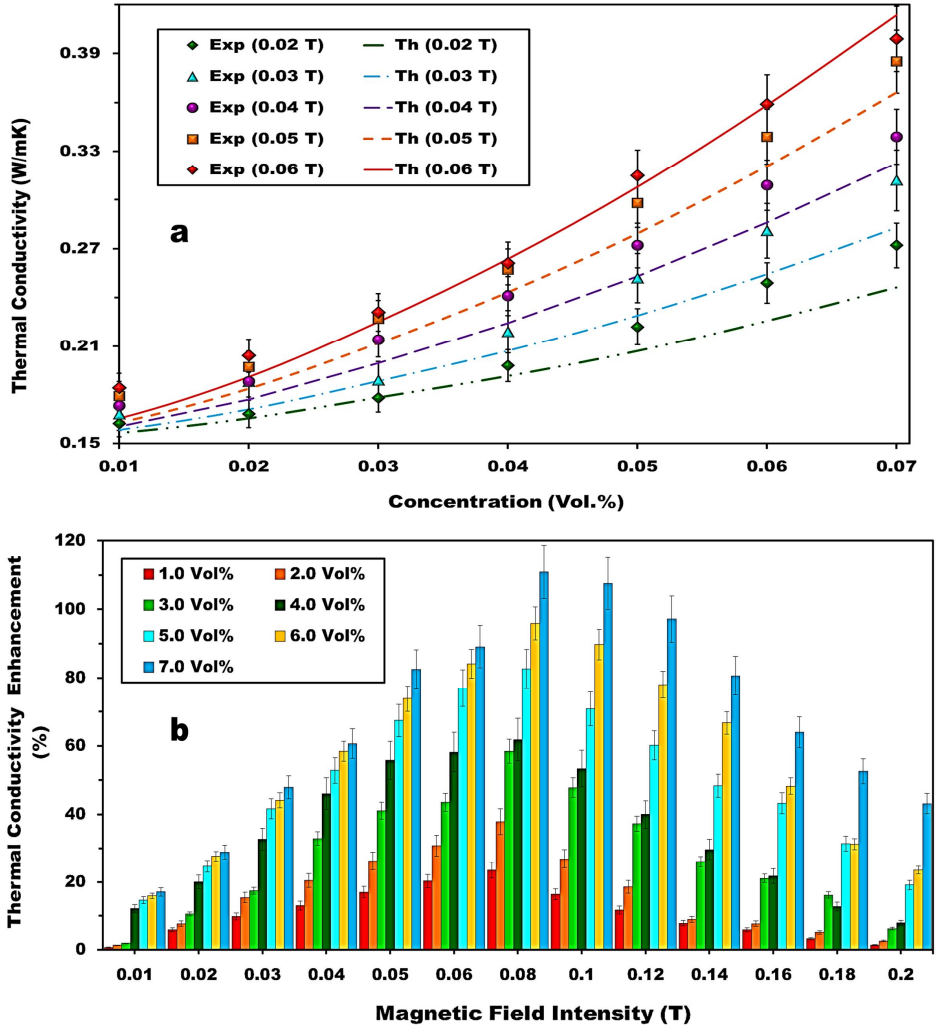

**Figure 5:** (a) Thermal conductivity enhancements of $Fe_3O_4$/HTO as a function of NP concentration at different magnetic field intensities and validation of the proposed model with experimental data. (b) Magnitude of corresponding enhancement of thermal conductivity as a function of magnetic field at different concentration of NPs.

When the energy of the system becomes sufficient enough to overcome the potential barrier limiting the lateral aggregation, lateral coalesce of the chains takes place. Beyond the CMF, the mean chain separation, 'd$_{sep}$' decreases drastically due to maximum chain density, which results in the energy overshooting the barrier. As the chains begin to aggregate to reduce the energy state of the system, the effective concentration of the fluid phase localized near the chains enhances, causing the thermal conductivity to decay beyond the CMF. As observed from the MH curves, the Fe$_2$O$_3$ NPs have low magnetic saturation compared to the Fe$_3$O$_4$ NPs at same magnetic field intensities. Therefore, zipping of chains for the Fe$_3$O$_4$ MNCs occur at higher magnetic fields compared to the Fe$_2$O$_3$ counterparts. Images of chain formation by nanoscale particles under magnetic field cannot be captured by normal imaging techniques; however, large scale zippering at very high field intensities beyond the CMF can be captured by large zoom photography. Large scale zippering photographed at high optical zoom, leading to visible separation of the agglomerates and the fluid phase in Fe$_2$O$_3$/kerosene and Fe$_3$O$_4$/HTO MNCs have been illustrated in Fig. 7 (c2) and (d2) respectively.

Although the physics and mechanism of enhanced thermal conduction can be established based on the principle of field induced thermally stable fibrillation, predictive mathematical tools to determine the magnitude of transport properties based on system information is required for design of such smart MNCs. An analytical model has been proposed along the lines of the physical mechanisms discussed so as to predict the augmented thermal conductivity in presence of magnetic fields. Experimentally, it has been observed that the thermal conductivity of the MNCs exhibits a second order polynomial trend with respect to the applied magnetic field intensity. Accordingly, the base form of the equation relating the effective thermal conductivity ($k_{eff}$) and magnetic field intensity ($B$) can be expressed as a simplistic quadratic as

$$k_{eff} = a_1 B^2 + a_2 B + a_3 \qquad (3)$$

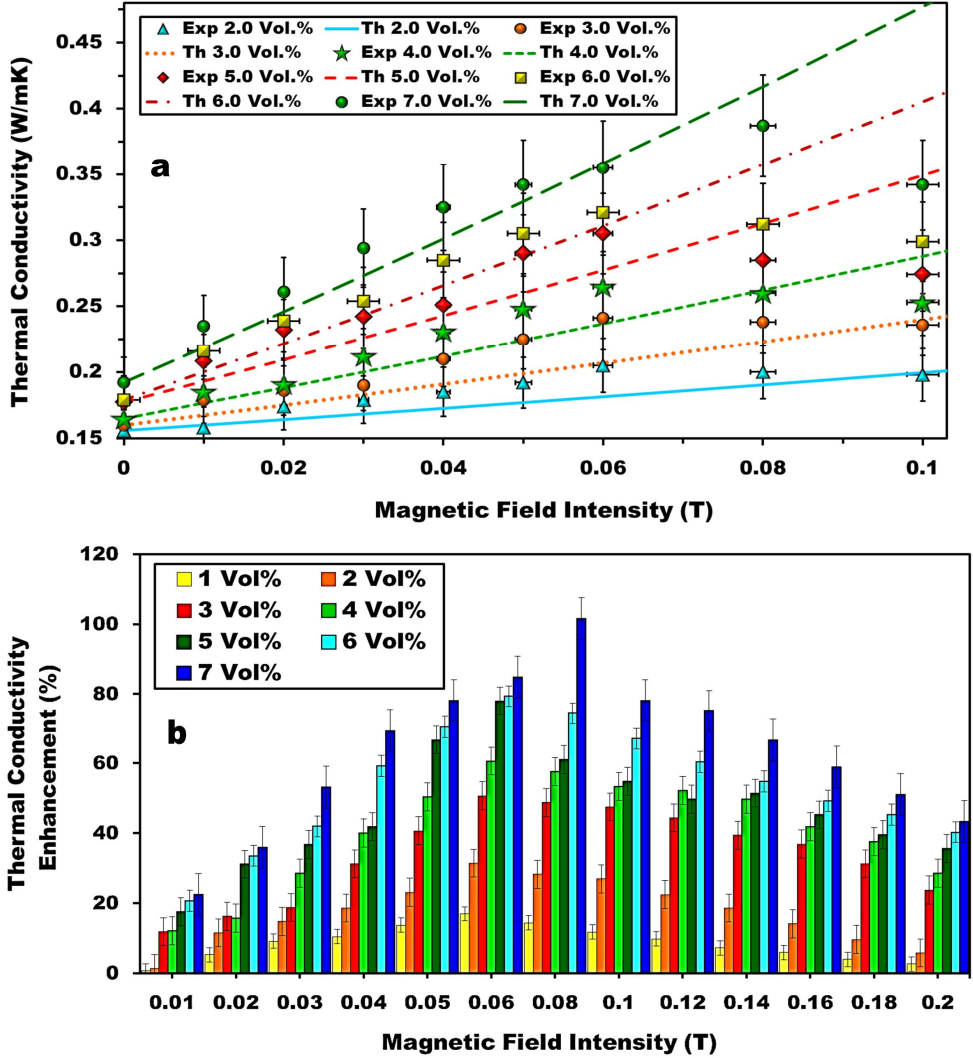

**Figure 6:** Thermal conductivity enhancement of $Fe_3O_4$/kerosene MNCs as a function of magnetic field intensity, (a) demonstrating the magnitude of thermal conductivity as well as validation of the proposed analytical model. (b) Illustrates the magnitudes of corresponding enhancement of thermal conductivity at different NP concentration.

The coefficients, $a_1$, $a_2$ and $a_3$, require being modeled based on system properties in such a way that Eqn. 3 is dimensionally consistent. Based on dimensional analysis, $a_3$ is a term equivalent to the thermal conductivity in the absence magnetic field, and thus is the thermal conductivity of the base MNC ($K_{nf,wf}$). Eqn. 3 therefore can be expressed in its new form as

$$k_{eff} = a_1 B^2 + a_2 B + k_{nf,wf} \qquad (4)$$

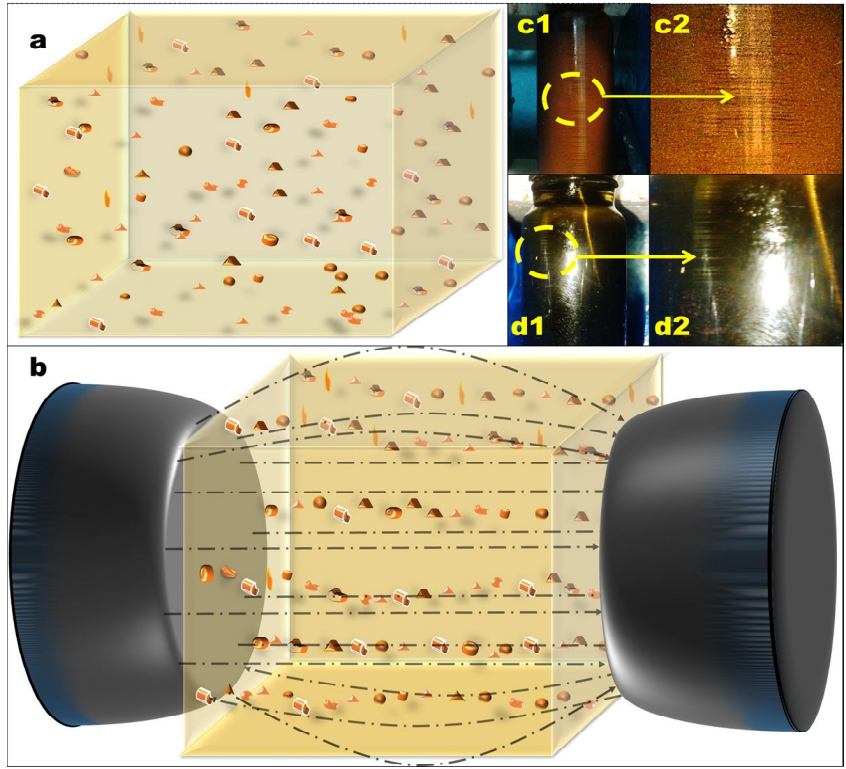

**Figure 7:** Illustration of the phenomena of (a) randomly scattered NPs in the absence of magnetic field. (b) Chain formation under the influence of external magnetic field. (c1) $Fe_2O_3$/kerosene MNC in the presence of magnetic field much higher than the CMF large scale zippering with visible localized segregation of the solid and the fluid phases (c2) Enlarged view of zippering phenomena in $Fe_2O_3$ NPs in the suspension. (d1) Similar phenomena in $Fe_3O_4$/HTO MNCs (d2) Enlarged view of zippering in $Fe_3O_4$/HTO MNCs.

The enhancement in thermal conductivity of the MNCs can be explained based on the interaction among the dispersed nanoparticles and how the behavior changes in presence of a magnetic field. One of the significant parameters which govern the magnetic behavior of the nanomaterials is the magnetic saturation ($H_{sat}$) and is the field strength at which the dispersed particles attain magnetic saturation. Thus '$H_{sat}$', essentially provides a barrier on the chain formation propensity of the particles as the saturation point demarcates the maximum competence of the particles to form chains and to participate in the enhancement of thermal conductivity. The dimensional analysis reveals that the coefficient '$a_1$' requires specific length and time scales for mathematical rendition. For MNCs, the most significant time scale that plays a direct role is the Neel-Brownian relaxation time scale [35]. At finite temperatures, the magnetization has the probability to flip and reverse its direction. Therefore, the attempt time period and its interaction with the thermal fluctuation of the particle directly determine the ultimate strength of the field induced chained structure. If the attempt period [36] is short, the more is the chance of strong chained structure formation due to longer duration of unhampered magnetic force of attraction between the individual particles. The coefficient '$a_1$' can be expressed mathematically in terms of magnetic saturation ($M_{sat}$), MFI critical chain length ($L_c$), absolute temperature ($T$), attempt period ($\tau_0$) of the NPs and the average number of particles required such that the chain is just thermally stable ($N_c$), as

$$a_1 = \left( \frac{M_{sat} L_c^2 N_c^\alpha \phi^n}{T \tau_0} \right) \tag{5}$$

The exponents '$n$' and '$\alpha$' in Eqn. 5 have been observed to be consistently possessive of the magnitudes varying from 1.5 – 2. The chain length influences the number density of chains in the colloids, i.e. as the chains length increases, the number density of the chains in the colloid reduces and hence, the thermal conductivity reduces. Thereby, the chain length should be optimized so as to be thermally stable, yet not long enough so as to reduce the number density. Since chain length within colloids is a statistical parameter that cannot be modeled analytically, the minimal value of chain length ($L_c$), such that chains shorter than this cannot exist due to thermal fluctuations, have been utilized as the worst-scenario condition. Mathematically, the minimal chain length of the NPs, determined from the concept of thermally stable chain structure [37, 38] can be expressed from the Brownian diffusivity of the chain and their settling velocity (obtained from the Stoke-Eisenstein equation) as

$$L_c \geq \left( \frac{12 k_B T}{\pi (\rho_p - \rho_f) g} \right)^{1/4} \tag{6}$$

Where '$T$', $k_B$ '$\rho_p$' and '$\rho_f$' are the absolute temperature, Boltzmann constant, density of particulate phase and density of the base fluid phase respectively. Eqn. (6) was found to accurate within the ±20 % of the results obtained through numerical simulations [39].

Accordingly, as reported in details in the previous report by the authors [28], the mathematical expression for the coefficient '$a_2$' is expressible as

$$a_2 = \left( \frac{L_c \mu_f N_c^{\beta} \phi^n}{\mu_0 \pi d_p \rho_f T} \right) \quad (7)$$

Where '$\mu_f$', '$\mu_0$' and '$d_p$' is indicate the dynamic viscosity of the base fluid, the magnetic permeability of free space and diameter of the NPs respectively. The exponent '$\beta$' depends on the magnitude of magnetic saturation of NPs and has been observed to be consistently possessive of values within the range of 2 – 3. The proposed semi-analytical model has been validated with respect to experimental data and good accuracy has been observed. Since the present model does not account for the zipping phenomena at high fields, it is only valid up to the CMF and cannot predict the deteriorated transport beyond the same. Essentially, the model acts as a predictive tool for the maximum possible thermal performance exploitable from a MNC under the effect of magnetic field.

## 4.2 $Fe_2O_3$ MNCs

The magnitudes of field induced enhanced thermal conductivity as well as the corresponding enhancements for $Fe_2O_3$/EG, $Fe_2O_3$/HTO, $Fe_2O_3$/kerosene MNCs as functions of magnetic field intensity have been illustrated in Figs. 8, 9 and 10 respectively. The maximum rise in the magnitude of thermal conductivity for each set of MNC is observed at CMF and as expected; the thermal conductivity drops beyond the CMF. The magnitude of maximum thermal conductivity for $Fe_2O_3$/EG MNCs is observed to be 0.512 W/mK (Fig. 8 (a)) for a 7 vol. % sample and the corresponding enhancement for the same with respect to base colloid is ~76 % (Fig. 8 (b)) at 0.12 T. The magnitude of maximum thermal conductivity for $Fe_2O_3$/HTO MNC is observed to be ~0.365 W/mK and the enhancement for the same is ~98 % at 0.1 T. For the $Fe_2O_3$/kerosene MNCs, these are ~0.307 W/mK (Fig. 9 (b)) and ~ 62 % (Fig. 9 (b)) at 0.06 T. However, the magnitudes of maximum enhancement in case of Fe2O3 based MNCs are lower than the corresponding $Fe_3O_4$ based MNCs due to the lower value of saturation moment of $Fe_2O_3$ NPs. It has been observed that the proposed model accurately predicts the phenomenon of thermal conductivity enhancement of $Fe_2O_3$ MNCs and shows good agreement with the present experimental results, as illustrated in Figs. 7(a), 8(a) and 9 (a).

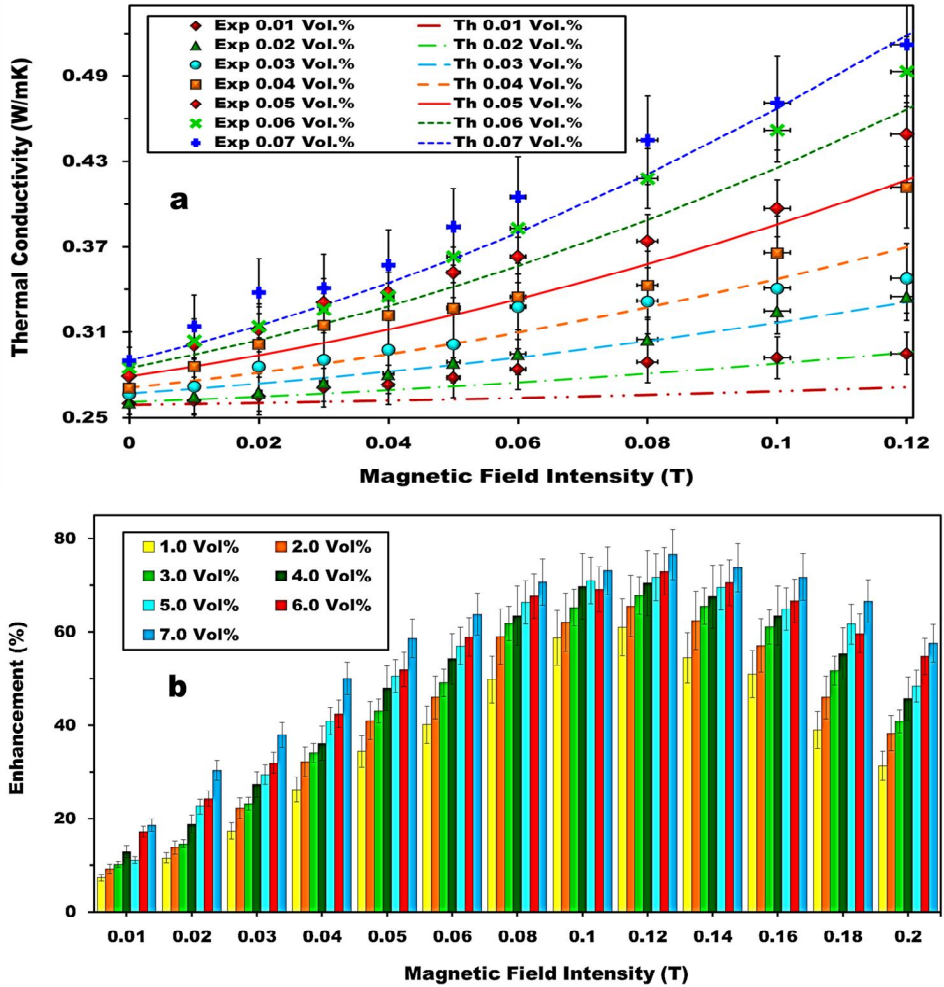

**Figure 8:** Thermal conductivity enhancement of $Fe_2O_3$/EG as a function of magnetic field intensity. (a) The magnitude of thermal conductivity and validation of the proposed model with experimental results. (b) The magnitude of respective enhancement of thermal conductivity at different NPs concentration.

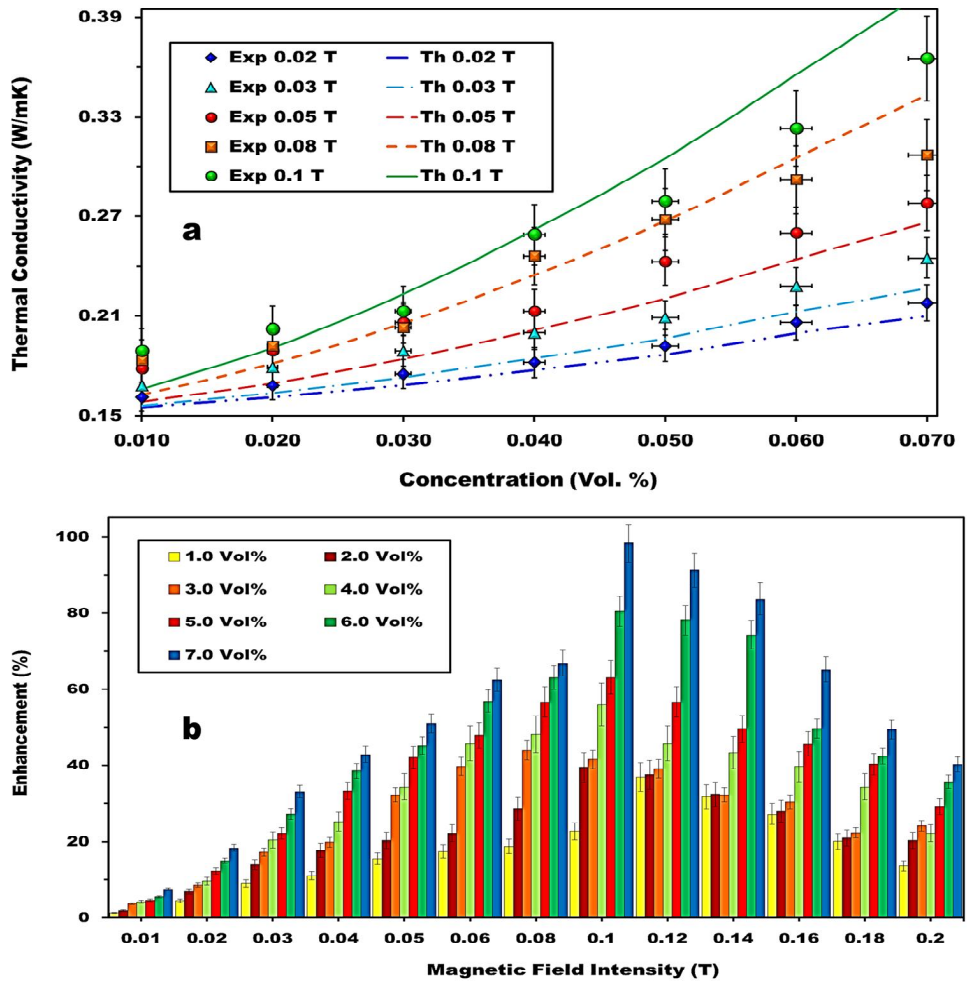

**Figure 9:** Thermal conductivity enhancements of $Fe_2O_3$/HTO MNCs as a function of magnetic field intensity. (a) The magnitude of thermal conductivity and validation of the model. (b) Illustrates the magnitude of respective enhancement of thermal conductivity at different $Fe_2O_3$ NPs concentration.

### 4.3 NiO and $Co_3O_4$ MNCs

Experiments for thermal conductivity of NiO and $Co_3O_4$ MNCs under magnetic field are conducted utilizing EG and kerosene as base fluids. The maximum magnitude of thermal conductivity for the samples of NiO/EG and NiO/Kerosene MNCs with 7 vol. % concentration of NPs is ~ 0.368 W/mK at 0.14 T (Fig. 11 (a1)) and 0.221 W/mK at 0.1 T (Fig. 11 (b1)) respectively, with the corresponding enhancements for the same being ~ 26 % and 29 % respectively. The maximum magnitude of the thermal conductivity enhancement for $Co_3O_4$/EG and $Co_3O_4$/kerosene MNCs is observed as ~ 4.0 % at 0.14 T (Fig 12 (a)) and ~ 3.4 % at 0.12 T (Fig. 12 ((b)) respectively. Since the saturation moment of NiO and $Co_3O_4$ NPs is relatively low compared to the iron oxide NPs, much reduced augmentation in thermal conductivity is observed the colloids. The present parameterization clearly reveals that from the point of view of utility, iron (II, III) oxide and iron (III) oxide nanostructure based MNCs are the best candidates for designing such colloidal systems whereas EG proves to be the best base fluid as far as stability and enhanced transport capabilities are concerned.

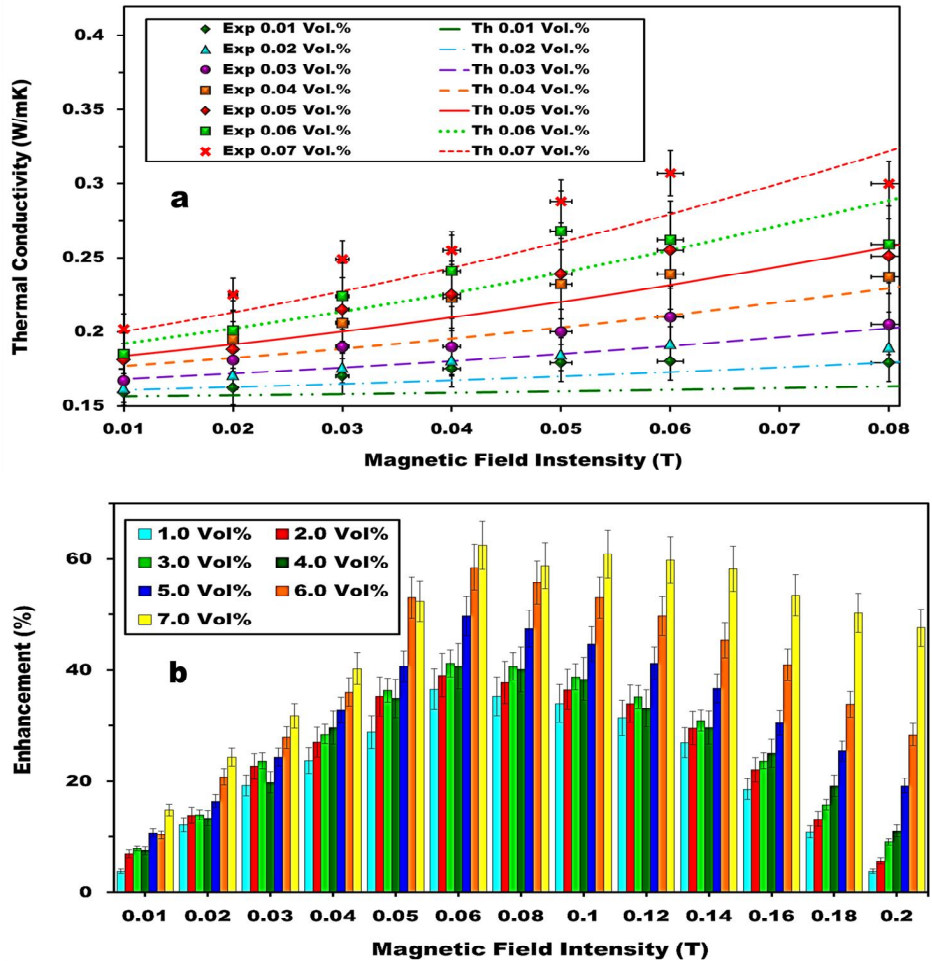

**Figure 10:** Thermal conductivity enhancements of $Fe_2O_3$/kerosene MNCs as a function of magnetic field intensity. (a) The magnitude of thermal conductivity and validation of the analytical model. (b) The magnitudes of enhancement of thermal conductivity at different $Fe_2O_3$ NP concentrations.

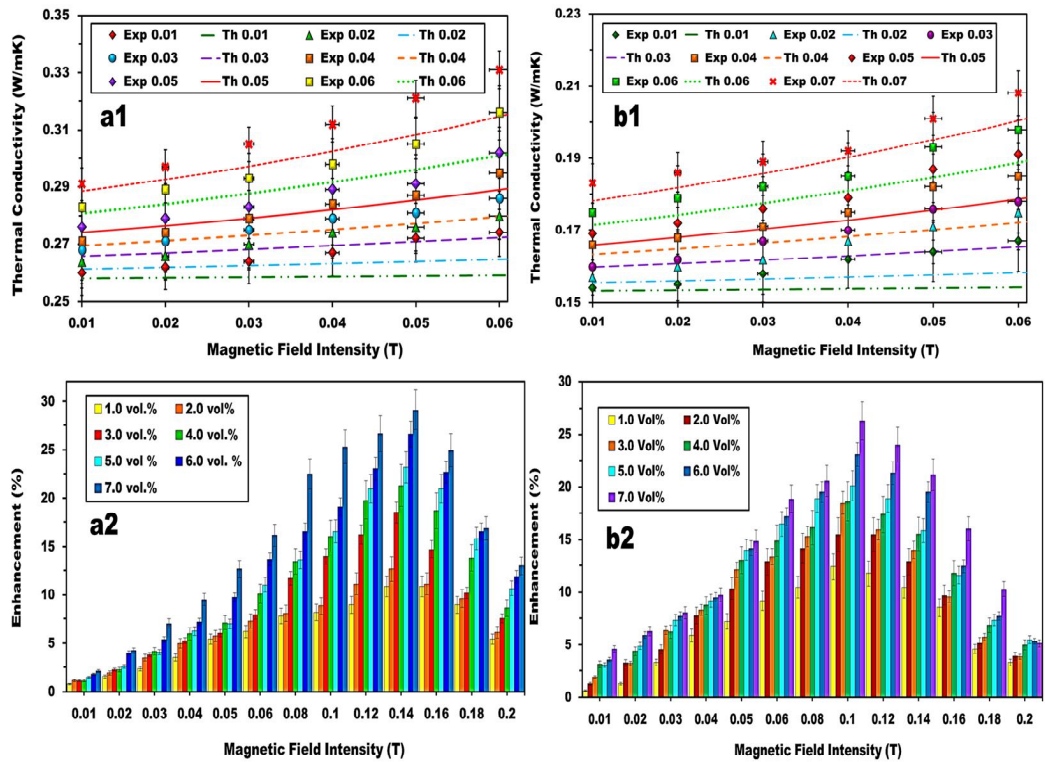

**Figure 11:** Thermal conductivity enhancement of NiO MNCs as a function of magnetic field intensity (a1) NiO/EG MNCs show maximum thermal conductivity at 0.14 T. (b1) NiO/kerosene MNCs show maximum thermal conductivity at 0.08 T. (a2) NiO/EG based MNCs yield maximum augmentation of ~30 % whereas (b2) NiO/kerosene yield ~ 25 %.

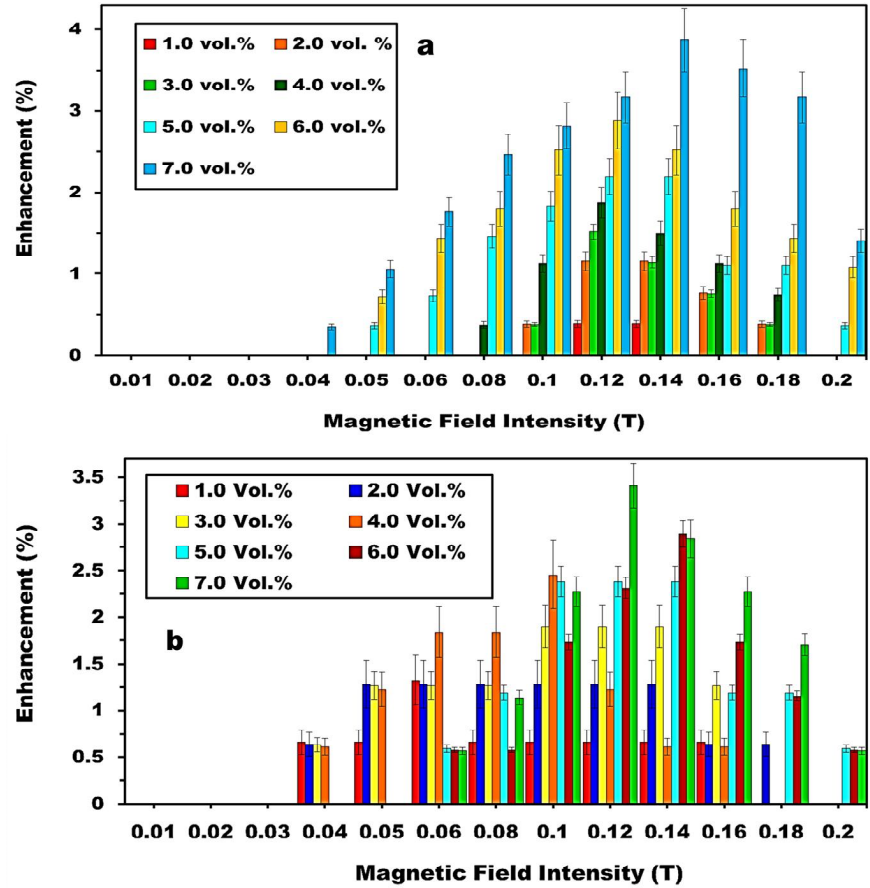

**Figure 12:** Thermal conductivity enhancement of $Co_3O_4$ MNCs as a function of magnetic field intensity. (a) Maximum magnitude of thermal conductivity enhancement of $Co_3O_4$/EG MNCs is ~ 4.0 % (b) $Co_3O_4$/kerosene MNCs show maximum augmentation of ~ 3.5 %.

## 5. Conclusions

Magnetic field induced augmented thermal conductivity of MNCs has been reported in the present article. A maximum augmentation in thermal conductivity of ~114 % has been observed for $Fe_3O_4$/EG MNC at 0.1 T for 7.0 vol. % concentration. It is also observed that $Fe_2O_3$, NiO and $Co_3O_4$ based MNCs have less thermal conductivity enhancement compared to $Fe_3O_4$ based MNCs at the same magnetic field and concentration due to lower values of saturation moment of the former. However, the performance of $Fe_2O_3$ based MNCs is appreciably good as far as thermal transport is concerned. It is observed that the CMF shifts towards the lower field intensities magnitude as the base fluid viscosity reduces and this has been explained on the basis of fibrillation characteristics of nanoparticles within the fluid medium in question. Further, it is observed that the transport capabilities decay out beyond the CMF and this has been explained based on zippering phenomena of the fibrils. A semi-analytical model has been proposed to correlate the physics and mechanism of the phenomena at play so as to predict the enhancement of thermal conductivity of MNCs based on system parameters. The model has been observed to accurately predict the experimental observations. The article thus provides a comprehensive approach towards understanding magnetic field induced enhanced thermal transport in magnetic nanofluids and design parameters for utility of such colloids in thermal management systems where magnetic fields can be produced at ease to enhance conductive cooling, such as, high power density drives, locomotives, electrical power systems, etc.

## Acknowledgements

The authors are thankful to the Sophisticated Analytical Instruments Facility (SAIF), IIT Madras for material characterizations. AK is also thankful to Dr. V. Ramanujachari, Director, Research and Innovation Centre (DRDO), Chennai, for fruitful technical discussions and for permission to publish the present research. Authors acknowledge the Defence Research and Development Organization (DRDO) of India for funding of the present work (Grant no. ERIP/ER/RIC/2013/M/ 01/2194/D (R&D)). PD also thanks the Ministry of Human Resource and Development (MHRD), Govt. of India, for the doctoral research scholarship.